\documentclass[sigconf, nonacm]{acmart}

\usepackage{array}
\newcolumntype{P}[1]{>{\raggedright\arraybackslash}p{#1}}

\usepackage{orcidlink}

\AtBeginDocument{%
  }

\setcopyright{acmlicensed}
\copyrightyear{2025}
\acmYear{2025}
\acmDOI{XXXXXXX.XXXXXXX}

\acmConference[HT '25]{36th ACM Conference on Hypertext and Social Media}{September 15--19, 2025}{Chicago, IL, USA}
\acmBooktitle{HT '25: Proceedings of the 36th ACM Conference on Hypertext and Social Media (HT '25), September 15--19, 2025, Chicago, IL, USA}
\acmISBN{978-1-4503-XXXX-X/25/09}

\graphicspath{{./images/}}

\begin{document}

\title{Sentiment and Social Signals in the Climate Crisis: A Survey on Analyzing Social Media Responses to Extreme Weather Events}

\author{Pouya Shaeri}
\email{pshaeri@asu.edu}
\orcid{0009-0009-5139-7654}
\affiliation{%
  \institution{School of Computing and Augmented Intelligence, Arizona State University}
  \city{Tempe}
  \state{Arizona}
  \country{USA}
}
\authornote{Pouya Shaeri and Yasaman Mohammadpour contributed equally to this work.}

\author{Yasaman Mohammadpour}
\email{ymoham15@asu.edu}
\orcid{0009-0005-0689-3123}
\affiliation{%
  \institution{New College of Interdisciplinary Arts and Sciences, Arizona State University}
  \city{Tempe}
  \state{Arizona}
  \country{USA}
}
\authornotemark[1]

\author{Alimohammad Beigi}
\email{abeigi@asu.edu}
\orcid{0009-0009-6637-0761}
\affiliation{%
  \institution{School of Computing and Augmented Intelligence, Arizona State University}
  \city{Tempe}
  \state{Arizona}
  \country{USA}
}

\author{Ariane Middel}
\email{ariane.middel@asu.edu}
\orcid{0000-0002-1565-095X}
\affiliation{%
  \institution{The GAME School,\\ Arizona State University}
  \city{Tempe}
  \state{Arizona}
  \country{USA}
}

\renewcommand{\shortauthors}{Shaeri and Mohammadpour, et al.}

\begin{abstract}
 Extreme weather events driven by climate change, such as wildfires, floods, and heatwaves, prompt significant public reactions on social media platforms. Analyzing the sentiment expressed in these online discussions can offer valuable insights into public perception, inform policy decisions, and enhance emergency responses. Although sentiment analysis has been widely studied in various fields, its specific application to climate-induced events, particularly in real-time, high-impact situations like the 2025 Los Angeles forest fires, remains underexplored. In this survey, we thoroughly examine the methods, datasets, challenges, and ethical considerations related to sentiment analysis of social media content concerning weather and climate change events. We present a detailed taxonomy of approaches, ranging from lexicon-based and machine learning models to the latest strategies driven by large language models (LLMs). Additionally, we discuss data collection and annotation techniques, including weak supervision and real-time event tracking. Finally, we highlight several open problems, such as misinformation detection, multimodal sentiment extraction, and model alignment with human values. Our goal is to guide researchers and practitioners in effectively understanding sentiment during the climate crisis era.
\end{abstract}


\maketitle

\section{Introduction}
As climate instability intensifies, societies worldwide are grappling not only with the physical consequences of extreme weather events but also with their psychological and sociocultural impacts~\cite{gabric2023climate, singh2023comprehensive, mostafa2025interconnected}. Catastrophic events such as wildfires, hurricanes, floods, heatwaves, and droughts—driven by human-induced climate change are occurring with increasing frequency, severity, and unpredictability~\cite{seneviratne2024extreme, floranza2019impact, shi2018impact}. These events do not occur in isolation; they are part of a complex web of social narratives, political discourse, media framing, and public emotion~\cite{carvalho2009reporting, al2019social, liao2024decoding}. Understanding how individuals and communities emotionally and behaviorally react to such events is crucial for a wide range of stakeholders, including climate scientists, public health officials, policymakers, emergency responders, and communication strategists~\cite{mahmoodimpact, bergquist2019experiencing}.

One of the most direct and unfiltered sources of public response in the modern era is social media. Platforms like Twitter (now X), Reddit, Facebook, and TikTok have become essential spaces where people share their experiences, express their concerns, voice their anger or solidarity, and seek information or emotional support during and after climate-related disasters~\cite{mavrodieva2019role, zein2024social, azghan2023personalized, fernandez2016talking}. These digital platforms serve not only as outlets for individual sentiment but also as sociotechnical systems that influence collective understanding and public opinion~\cite{makela2024climate, pupneja2023understanding, mair2011events}. They offer a dynamic and rich source for computational sentiment analysis, allowing for the study of emotional responses to climate events on a large scale, in real-time, and across various demographic and geographic segments~\cite{sultana2024systematic}.

\begin{figure*}[ht]
  \centering
  \includegraphics[width=0.80\linewidth]{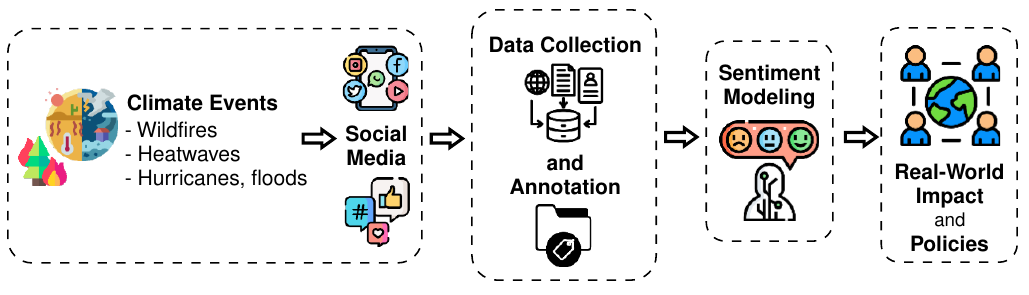}
  \caption{
    Overview of the climate sentiment pipeline linking extreme events to social media reactions, analyzed through data, models, and their societal impact. Our survey follows this structure, reviewing existing literature across each stage of the pipeline.
  }
  \Description{Diagram showing the flow from climate events to social media reactions, followed by stages such as data collection, modeling, and real-world applications like policy and health monitoring.}
  \label{fig:overview}
\end{figure*}

However, mining sentiment from social media during climate crises is a complex task. Unlike structured surveys or formal reports, social media data is unstructured, fleeting, and multimodal~\cite{kokoschka2024visual, vivion2024misinformation, shaeri2025multimodalphysicsinformedneuralnetwork, al2022covid}. Posts may include sarcasm, coded language, memes, images, and contextual references that require deep linguistic and cultural understanding for accurate interpretation~\cite{makela2024climate, pranto2024satire}. Additionally, the emotional tone of climate discourse is often intertwined with political ideology, narratives of environmental justice, misinformation, and conspiracy theories, which complicates efforts for objective analysis~\cite{gimello2025embers, webb2016digital, lakzaei2025neighborhood}. The rapid changes in trending hashtags, new slang, and platform-specific conventions contribute to the linguistic volatility of platforms, making it more challenging~\cite{al2019social, pignot2023affective}.

Despite these challenges, sentiment analysis in this field shows great potential. By examining how individuals feel, what they fear, whom they blame, and which solutions they support or reject, researchers can gain a better understanding of the emotional context surrounding climate events~\cite{liao2024decoding, pupneja2023understanding, ko2024experience}. This can lead to better empathetic disaster communication strategies, identify mental health stress signals, reveal misinformation dynamics, and guide policy discussions~\cite{bracchetti2024mass, thomasimpacts, dey2021natural, adamopoulos2024climate}.

In addition to sudden climate disasters like wildfires and hurricanes, chronic stressors such as extreme summer heat in desert cities are becoming an increasing public health concern. Urban areas in the U.S. Southwest, in particular, face heightened warming due to the Urban Heat Island (UHI) effect, which can often exceed tolerable limits during summer heatwaves~\cite{middel2023urban, schneider2024disconnect, alkhaled2024webmrt}. Recent studies indicate that prolonged exposure to heat can result in heat stress, physiological strain, and emotional exhaustion, particularly among vulnerable groups such as unhoused individuals~\cite{karanja2024impact, shakib2024mapswipe}. Discussions about these impacts are increasingly prevalent on social media, where public sentiment reflects distress and frustration regarding inadequate urban infrastructure. There are also rising calls for cooling interventions, such as shade or misting systems~\cite{huang2025outdoor, browning2024value, thakar2021techno}. By considering extreme heat as a slow-onset climate disaster, we broaden the focus of this survey to incorporate the emotional and communicative aspects of prolonged thermal crises.

\footnotetext[1]{The full and extended literature review—including all references analyzed and categorized beyond those cited in this manuscript—is available at the project’s GitHub repository: \url{https://github.com/pouyashaeri/ClimateSentiment}}

This survey provides a thorough overview of sentiment analysis techniques used in social media related to weather and climate change events~\footnotemark[1]. While there are general reviews available, very few specifically address the unique challenges, methodologies, and insights that emerge at the intersection of climate disasters and public sentiment~\cite{vivion2024misinformation, ahmad2022climate, sayigh2023global}. To frame our discussion, we will use the 2025 Los Angeles forest fires as a case study. This large-scale wildfire sparked intense online discourse, characterized by grief, outrage, political polarization, and systemic critique~\cite{farahmand2025attengluco, gimello2025embers}.

We start by discussing the fundamental concepts in sentiment and emotion analysis. We categorize the approaches into several types, including early lexicon-based methods, traditional machine learning models, and more recent deep learning architectures. Additionally, we highlight the emerging use of large language models (LLMs) like GPT-4 and LLaMA~\cite{beigi2024lrq, li2024generation, lakzaei2025decision}.

We examine methods for real-time data collection and annotation, including hashtag tracking, social media scraping, and weak supervision, which are used to create labeled datasets during rapidly evolving climate events~\cite{olteanu2015comparing, al2019social, fernandez2016talking}. Special attention is given to the challenges of annotating sarcasm, emotional nuance, and cross-cultural expressions of emotion in high-stakes and politically sensitive contexts~\cite{makela2024climate, pranto2024satire, pupneja2023understanding}. We also examine current evaluation strategies and datasets, highlighting key research gaps: the absence of multilingual and multimodal corpora, insufficient real-time modeling of sentiment evolution, and the inadequate representation of misinformation-aware sentiment systems~\cite{kokoschka2024visual, yosipof2024cyber, vivion2024misinformation, lakzaei2024disinformation}. A primary motivation for this survey is the increasing prevalence of AI-generated misinformation during climate emergencies.~\cite{yosipof2024cyber, gimello2025embers, michail2024incorporating}. Events like the 2025 LA forest fires revealed how synthetic images and videos, such as fake depictions of burning Hollywood landmarks, can mislead the public~\cite{chappell2025fakeimages, Afshar2025, gimellodecoding}.

In addition to methodological synthesis, we explore how sentiment analysis is incorporated into climate adaptation and mitigation frameworks. This includes its use in crisis dashboards, policy responsiveness, and early warning systems for emotional well-being~\cite{hart2024climate, garcia2024wildfires, bergquist2019experiencing, fernandez2016talking}. We also emphasize efforts that extend beyond classification to include psychological aspects of emotional trauma, distress, and collective feelings during environmental crises~\cite{metzen2024psychology, lasoff2024supporting, thoma2021clinical}. Finally, we consider the broader societal implications, including concerns about surveillance, algorithmic bias, and the potential misuse of affective computing, especially in ways that may reinforce inequality or marginalize vulnerable voices~\cite{makela2024climate, webb2016digital, lakzaei2025neighborhood, alsadat2024distributed, alkhaled2024webmrt}. This paper outlines a roadmap for interdisciplinary collaboration among NLP, social computing, environmental communication, and digital humanities, demonstrating how emotional signals in digital discourse can enhance our understanding and response to the climate crisis (Figure ~\ref{fig:overview}).

\vspace{-0.15cm}
\section{Preliminaries}

This section defines key terms, techniques, and modeling paradigms used throughout the discussion. The preliminaries orient readers from diverse disciplines—ranging from NLP and machine learning to climate science and social media studies—by introducing both domain-specific terminology and the computational tools applied to sentiment analysis in the context of climate-related social discourse.

\vspace{-0.15cm}
\subsection{Key Concepts}

\noindent\textbf{Sentiment Analysis}: Also known as opinion mining, sentiment analysis is the computational task of identifying and categorizing the emotional valence expressed in a given piece of text. Most commonly, this involves labeling user-generated content—such as tweets, Reddit posts, or Facebook comments—as \textit{positive}, \textit{negative}, or \textit{neutral}. In the context of climate-related events, sentiment analysis helps detect expressions of outrage, fear, solidarity, blame, or hope in the aftermath or buildup of environmental crises~\cite{liu2010sentiment, Beigi2016}.

\noindent\textbf{Emotion Detection}: Emotion detection extends beyond binary or ternary sentiment classification by identifying nuanced emotional states such as anger, anxiety, empathy, sadness, joy, or frustration. These emotional signals provide a finer-grained lens into public perception and can be particularly important during climate disasters where emotional polarity alone may not be sufficient to capture the full spectrum of human reactions~\cite{garcia2024wildfires, pupneja2023understanding}. Emotion detection models often rely on specialized emotion lexicons or are fine-tuned using emotion-labeled corpora such as GoEmotions~\cite{demszky2020goemotions} or EmpatheticDialogues~\cite{rashkin2018towards}.

\noindent\textbf{Heat Stress}: Heat stress refers to the physiological and psychological strain that occurs when the human body is exposed to extreme heat, particularly when it cannot effectively regulate its internal temperature. This condition is exacerbated in urban areas due to the Urban Heat Island (UHI) effect, where built environments retain heat more than natural landscapes. In desert cities like Phoenix or Las Vegas, heat stress can reach crisis levels during summer, posing severe health risks to vulnerable populations including outdoor workers, children, the elderly, and unhoused individuals~\cite{guzman2024development, hart2024climate, turner2022cities}. In the context of climate communication, heat stress is not only a public health issue but also a recurring emotional theme in social media posts, where users express anxiety, exhaustion, and anger about ongoing exposure and insufficient urban adaptation~\cite{middel2019micrometeorological, bergquist2019experiencing}.

\noindent\textbf{Event-Centric Sentiment}: Unlike general sentiment analysis, event-centric sentiment focuses on linking emotional or affective expressions to specific real-world events or triggers. For instance, a tweet expressing outrage about delayed wildfire evacuation efforts may be tagged not only as negative but also temporally and contextually tied to a specific climate event (e.g., the 2025 Los Angeles forest burns)~\cite{ko2024experience, sadik2022analyzing}. This approach enables sentiment tracking across the timeline of an event and supports temporal and causal modeling of emotional trajectories.

\noindent\textbf{Misinformation and Disinformation}: In climate discourse, the line between fact and fiction can often be blurred. \textit{Misinformation} refers to false or misleading information spread without harmful intent, whereas \textit{disinformation} denotes the deliberate spread of such content to manipulate or deceive~\cite{vivion2024misinformation, pranto2023bad}. Both phenomena can significantly distort sentiment analysis outcomes by influencing public perception and emotional response to climate events. For example, false narratives about the causes of a wildfire or conspiracy theories surrounding geoengineering may amplify anger or fear within online communities~\cite{gimello2025embers}.

\vspace{-0.3cm}

\subsection{Modeling Paradigms}

The sentiment analysis pipeline generally involves preprocessing social media data, extracting features, training models, and evaluating predictions. A wide range of computational models have been employed over the years, and their capabilities have evolved alongside advances in machine learning and NLP:

\noindent\textbf{Traditional Machine Learning Models}: Early approaches to sentiment analysis often relied on hand-engineered features—such as n-grams, TF-IDF scores, and part-of-speech tags—fed into classical classifiers including Support Vector Machines (SVMs), Logistic Regression, Decision Trees, and Random Forests. Although these models remain interpretable and computationally efficient, they struggle with informal language, sarcasm, and domain-specific terminology prevalent in climate discourse on social media~\cite{sadik2022analyzing, fernandez2016talking}.

\noindent\textbf{Neural Models}: With the rise of deep learning, neural architectures like Convolutional Neural Networks (CNNs) and Long Short-Term Memory networks (LSTMs)~\cite{staudemeyer2019understanding} gained prominence for sentiment analysis. These models are able to capture sequential dependencies and local context in text data, better handling noisy and informal social media language. In climate applications, LSTM-based models have been applied to wildfire discourse to analyze temporal emotional patterns \cite{garcia2024wildfires}, while CNNs were employed in visual-text sentiment fusion models in disaster-related image captioning \cite{michail2024incorporating}.

\noindent\textbf{Transformer-Based Models}: The introduction of transformer architectures, particularly BERT (Bidirectional Encoder Representations from Transformers)~\cite{devlin2019bert} and its variants (e.g., RoBERTa~\cite{liu2019roberta}, DistilBERT~\cite{sanh2019distilbert}, ClimateBERT~\cite{webersinke2021climatebert}), has revolutionized text classification tasks. Transformers are pre-trained on massive corpora and fine-tuned for downstream tasks with relatively small labeled datasets. They excel at capturing long-range dependencies and contextual semantics~\cite{pupneja2023understanding, vivion2024misinformation}. In climate sentiment research, fine-tuned transformers have been shown to outperform traditional models in handling domain shifts, contextual nuance, and multi-lingual inputs~\cite{shaeri2023semi, yosipof2024cyber}.

\noindent\textbf{Large Language Models (LLMs)}: The most recent advancement in the field comes with the emergence of generative large language models like GPT-3.5, GPT-4, Claude, and LLaMA. These models, trained on trillions of tokens, possess strong capabilities in zero-shot and few-shot learning through prompting. LLMs can be used not only for direct sentiment classification but also for generating rationales, explanations, and even synthetic labeled datasets ~\cite{tan2024large}. In the context of climate event sentiment analysis, LLMs offer the potential for real-time deployment and on-demand adaptation to new events without extensive retraining~\cite{li2024generation, ghatora2024sentiment}.

Together, these concepts and modeling paradigms establish the technical and theoretical basis for the rest of this survey. As we explore specific methodologies, datasets, and challenges in subsequent sections, we will refer back to these foundational ideas to situate each contribution within the broader landscape of sentiment analysis for climate communication.

\vspace{-0.2cm}
\section{Data Collection and Annotation}

The effectiveness of sentiment analysis for climate events depends largely on the quality of the analyzed data. Building accurate sentiment-aware models requires collecting representative, relevant, and high-fidelity social media data. However, this task demands careful strategies that consider platform differences, event-driven dynamics, and linguistic diversity~\cite{heo2022public, wang2024public}. Annotation quality is equally crucial, as it directly impacts model training, evaluation, and benchmarking. This section reviews major data sources and annotation methods in climate sentiment studies. Several works have leveraged hashtags, geolocation, and event-based keywords for social media data collection~\cite{al2019social, olteanu2015comparing}. Other studies have addressed challenges in annotating emotional nuances across cultures, especially during fast-evolving disaster contexts~\cite{makela2024climate, liao2024decoding}.

\vspace{-0.3cm}
\subsection{Sources of Climate-Related Social Media Data}

Social media provides a massive, real-time stream of user-generated content that can reflect the collective emotional landscape during climate events. Several platforms are particularly relevant for mining sentiment in this domain:

\noindent\textbf{Twitter/X and Mastodon}: Twitter remains the most widely used source for climate-related sentiment analysis due to its public nature, short-text format, and widespread use during emergencies. Researchers typically collect tweets using event specific hashtags (e.g., \#LAfire, \#ClimateCrisis, \#HeatWave), location-based filters, or keyword-based queries. Twitter’s real-time API or tools such as \texttt{Tweepy} and \texttt{Hydrator} are commonly used to scrape tweet content, metadata, timestamps, and geolocation (when available). Mastodon, a decentralized alternative to Twitter, offers similar microblogging functionality within a federated network of independent servers. Although less widely adopted, its public posts and chronological timelines can also serve as a source for event-based sentiment tracking~\cite{al2019social, ko2024experience, jeong2025fediversesharing}.

\noindent\textbf{Reddit}: Reddit hosts long-form discussions in climate-focused communities such as \texttt{r/climate}, \texttt{r/environment}, \texttt{r/news}, and \texttt{r/disasterpreppers}. Unlike Twitter, Reddit supports in-depth, context-rich exchanges through dialogue trees, debates, and nuanced emotional reflection. Data are commonly collected using Pushshift APIs or custom web scrapers to access posts and hierarchical comment threads for a sentiment analysis~\cite{makela2024climate}.


\noindent\textbf{News Article Comment Sections and Blogs}: Climate sentiment is also expressed in the comment sections of digital news articles, especially during high-visibility events. Publicly available data sets or scraped content from platforms such as The Guardian, CNN, or local newspapers provide complementary sources for sentiment extraction that differ in tone and vocabulary from traditional social media~\cite{olteanu2015comparing}.

\noindent\textbf{Multimodal Platforms (TikTok, Instagram, YouTube)}: Although this survey focuses on textual sentiment, it is important to note that platforms such as TikTok and Instagram contain rich emotional narratives shared through images, videos, captions, and comments. Although harder to process, these platforms offer a growing frontier for multimodal climate sentiment analysis~\cite{kokoschka2024visual, glodosky2024multimodal}.

\vspace{-0.3cm}
\subsection{Annotation Techniques for Climate Sentiment}

Once raw data is collected, it must be annotated with appropriate sentiment labels before it can be used for model training or evaluation. Climate-related social media data often contains subjective, ambiguous, or emotionally complex expressions that make labeling difficult. The following are the most commonly used annotation strategies in this domain:

\noindent\textbf{Manual Annotation}: Human annotators—either domain experts or crowdworkers recruited via platforms such as Amazon Mechanical Turk or Prolific—label social media posts based on predefined sentiment categories~\cite{d2022automatic, aslan2017human}. These may include basic polarity labels (positive, negative, neutral) or fine-grained emotions (e.g., anger, fear, hope, sadness). Manual annotation typically yields high-quality labels but is time-consuming, expensive, and often inconsistent across annotators, especially when the content is politically charged or emotionally subtle. Inter-annotator agreement (e.g., Cohen's kappa) is used to evaluate label reliability~\cite{liao2024decoding, alahmary2023semiautomatic}.

\noindent \textbf{Weak Supervision and Lexicon-Based Labeling}: Instead of relying on expensive manual labeling, many studies employ weak supervision techniques, including lexicon-based methods using resources such as VADER~\cite{hutto2014vader}, NRC Emotion Lexicon~\cite{mohammad2013nrc}, SentiWordNet\cite{baccianella2010sentiwordnet}, or AFINN \cite{desai2018sentiment}. These tools score words or phrases based on their emotional polarity or affective strength and aggregate sentiment across a post. While scalable, lexicon-based methods struggle with slang, sarcasm, negation, and domain-specific language (e.g., climate jargon or activist rhetoric), limiting their precision in climate sentiment applications~\cite{sultana2024systematic, fine2022language}.

\noindent \textbf{Rule-Based and Heuristic Methods}: In some cases, heuristic rules are defined based on hashtags, emojis, or common phrases~\cite{hutto2014vader}. For example, tweets containing \#ClimateEmergency or fire-related emojis may be assumed to convey negative sentiment. These techniques offer lightweight labeling strategies but must be used cautiously to avoid bias and overgeneralization~\cite{yilmaz2021multi, zhao2023toward}.

\noindent \textbf{LLM-Assisted Annotation}: The growing availability of powerful LLMs such as GPT-4 and Claude has enabled a new form of semi-automated annotation. With appropriate prompting, LLMs can label sentiment with context-awareness, explain their reasoning, and even provide probabilistic uncertainty estimates. In some studies, human annotators review or verify LLM-generated labels, creating hybrid annotation pipelines that combine scale with interpretability~\cite{beigi2024lrq, tan2024large}. However, LLM-generated labels may still reflect the model’s training bias or hallucinate interpretations, especially in emotionally charged or culturally sensitive contexts~\cite{ahadian2025survey, nahar2024fakes, beigi2024lrq}.

\noindent \textbf{Distant Supervision from Emojis or Reactions}: Some researchers use emojis or platform-specific reactions (e.g., Facebook’s “angry” or “sad” emojis) as distant proxies for emotional labels. While noisy, this method enables the creation of large-scale labeled corpora with minimal manual intervention and can be used for pre-training sentiment models~\cite{liu2021improving, al2022covid, shiha2017effects}.

\noindent \textbf{Annotation for Complex Tasks}: In addition to basic sentiment, some climate studies annotate related tasks such as stance detection (support/opposition to climate action), misinformation classification, emotional intensity, or cause attribution. These require task-specific guidelines and are often used to train multi-task models capable of deeper interpretation~\cite{emran2024masonperplexity, kim2022understanding, lakzaei2024disinformation, pranto2024satire}.

The annotation strategy selected influences not only model performance but also the type of insights that can be derived. Given the subjective and politicized nature of climate discourse, ensuring annotation consistency, transparency, and ethical oversight is critical for producing trustworthy and socially responsible sentiment analysis systems.

\vspace{-0.1cm}
\section{Methodology Taxonomy}

In this section, we present a structured taxonomy of methodologies used for sentiment analysis in the context of social media and climate-related events. These methods can be broadly grouped into five categories, reflecting both the historical evolution and current state-of-the-art in NLP: lexicon-based methods, traditional machine learning, deep learning architectures, transformer-based models, and LLMs. Each category brings unique strengths and limitations in handling the linguistic complexity, emotional nuance, and domain-specificity of climate discourse.
\vspace{-0.2cm}
\subsection{Lexicon-Based Methods}

Lexicon-based methods are among the earliest and most interpretable approaches to sentiment analysis. These rule-based models rely on predefined sentiment dictionaries—also known as lexicons—that associate individual words or phrases with sentiment scores. Prominent lexicons include:

\noindent\textbf{SentiWordNet}: An extension of WordNet that assigns polarity scores (positive, negative, objective) to synsets.

\noindent\textbf{VADER (Valence Aware Dictionary and sEntiment Reasoner)}: A lexicon and rule-based sentiment analysis tool designed specifically for social media text, accounting for capitalization, intensifiers, negations, and emoticons.

\noindent\textbf{NRC Emotion Lexicon}: Associates words with multiple emotional categories such as anger, anticipation, and joy.

\noindent\textbf{AFINN}: Assigns integer-based sentiment scores to words based on their emotional tone.

\noindent These methods are fast, transparent, and easy to implement, making them useful for exploratory analyses or as weak supervision signals in downstream tasks. However, they struggle with the informal, ambiguous, and context-dependent nature of social media language—particularly in climate-related discussions where sarcasm, irony, activism rhetoric, or technical terminology may not be represented in the lexicon. They are also limited in capturing phrase-level or sentence-level sentiment composition, where word-level polarity may be insufficient~\cite{sadik2022analyzing, ko2024experience, zamarreno2020social}.
\vspace{-0.1cm}
\subsection{Traditional Machine Learning Approaches}

Traditional supervised machine learning methods form the backbone of many early sentiment classifiers. These approaches typically involve transforming raw text into numerical feature representations—such as:
\textbf{Bag-of-Words (BoW)} (A simple frequency count of words in a document),
\textbf{Term Frequency-Inverse Document Frequency (TF-IDF)} (A weighting scheme that balances word frequency with distinctiveness across documents), and 
\textbf{N-grams and POS tags} (Features capturing short sequences or grammatical structures).

Once features are extracted, they are fed into classifiers such as
\textbf{Support Vector Machines (SVMs)},
\textbf{Naive Bayes},
\textbf{Logistic Regression}, and 
\textbf{Random Forests}.

While these models are relatively lightweight and interpretable, they rely heavily on feature engineering and cannot inherently model word order or semantics. This limits their understanding of nuanced sentiment or evolving slang in climate conversations. Nonetheless, they are still widely used as baselines or for ensemble methods when computational efficiency is prioritized~\cite{malviya2020machine, azghan2024cudle, alsadat2024multi}.

\subsection{Deep Learning Approaches}

Deep learning revolutionized sentiment analysis by enabling models to learn complex patterns in raw text data without requiring extensive manual feature engineering. The most common deep learning architectures include:

\textbf{Convolutional Neural Networks (CNNs)}: Originally designed for image processing, CNNs have been adapted for text classification by treating sentences as matrices of word embeddings. They are particularly good at capturing local n-gram patterns~\cite{kim2022case, kareem2021evaluation, farhangmehr2025spatiotemporal}.

  \textbf{Recurrent Neural Networks (RNNs)}: Including LSTM and Gated Recurrent Unit (GRU) variants, RNNs are designed to model sequential text dependencies and can capture the flow of sentiment in longer texts or threads~\cite{el2024comparative, gao2021novel, farhangmehr2025spatiotemporal}.

  \textbf{Attention Mechanisms}: These allow models to dynamically focus on relevant parts of the input text when making predictions, improving performance in emotionally complex or multi-topic inputs~\cite{li2024cllmate, marcondes2025case, shan2024correlation}.

Deep learning models outperform traditional approaches in capturing context, emotion, and non-linear relationships. However, they typically require larger labeled datasets and are less interpretable. Their performance may degrade when applied to domains (like climate discourse) with rare or emerging terms not well represented in pretraining data~\cite{garcia2024wildfires, azghan2025cudle, gimello2025embers}.

\subsection{Transformer-Based Models}

Transformer architectures, particularly since the introduction of BERT, have become the dominant paradigm in NLP. Transformers use self-attention mechanisms to model relationships between all tokens in a sentence simultaneously, allowing for deep contextual understanding recently in climate research.

\noindent Key transformer models used in sentiment analysis include:

\noindent \textbf{BERT (Bidirectional Encoder Representations from Transformers)}: A deeply bidirectional model pre-trained on masked language modeling and next sentence prediction~\cite{uthirapathy2023topic, kolbel2024ask, corringham2021bert}.

\noindent \textbf{RoBERTa}: A robustly optimized BERT variant trained with more data and dynamic masking~\cite{mohasina2023roberta, angin2022roberta}.

\noindent \textbf{DistilBERT and ALBERT}: Lightweight alternatives~\cite{george2024decoding, vaid2022towards}.

\noindent \textbf{ClimateBERT}: A domain-adapted BERT model fine-tuned on climate change-related texts, improving relevance for environmental discourse~\cite{webersinke2021climatebert, garrido2023fine, schimanski2023climatebert}.

These models can be fine-tuned on small labeled datasets for sentiment classification. They excel at handling context, sarcasm, negation, and multi-lingual sentiment, making them highly effective in climate-related social media analysis. However, their interpretability and computational cost remain areas of concern, especially for real-time deployment during rapidly unfolding events~\cite{klariza2023impact, azghan2025can}.

\subsection{LLMs for Sentiment Analysis}

The emergence of generative LLMs such as GPT-3.5, GPT-4, Claude, and LLaMA represents a new frontier in sentiment analysis. Unlike fixed classifiers, LLMs can be prompted dynamically to perform zero-shot or few-shot sentiment classification using natural language instructions~\cite{herrera2025overview, krugmann2024sentiment}. For example, a prompt such as:

\textit{"Determine the sentiment of the following tweet: `Another day of orange skies and toxic air. When will the fires end?'"}  
→ \textit{"Sentiment: Negative. Reason: The speaker expresses frustration and despair about ongoing wildfires."}

Large Language Models (LLMs) offer flexible sentiment analysis through prompt-based inference, enabling them to interpret sarcasm, moral nuance, and emotional intensity without task-specific fine-tuning~\cite{tanoenhancing, beigi2024lrq, bhargava2025impact}. They can also generate synthetic labeled data and rationales, supporting downstream models with improved training samples and richer emotional reasoning~\cite{sundarreson2024sentigen, huang2024targa}.

However, LLMs also have challenges: they may hallucinate labels or explanations, exhibit bias, or lack transparency in their reasoning. Their outputs often require human oversight, especially in sensitive or high-stakes domains such as climate misinformation or disaster response~\cite{waldo2024gpts, beigi2024model}.

The methodological landscape for sentiment analysis in climate discourse has evolved significantly over the last decade. The choice of method depends on the nature of the task, the quality and size of available data, the urgency of analysis, and the interpretability or ethical constraints of the application. In the following sections, we delve into the practical deployment of these models, supported by case studies and performance benchmarks. A comprehensive comparison of sentiment analysis approaches applied to climate-related social media is presented in Table~\ref{tab:methodology_comparison}, organizing key literature by methodology, performance, key findings, and limitations.

\begin{table*}[!htbp]
\centering
\caption{Comprehensive comparison of sentiment analysis methodologies for climate-related social media}
\label{tab:methodology_comparison}
\small
\begin{tabular}{P{3cm}P{2cm}P{2.6cm}P{3.2cm}P{4.1cm}}
\toprule
\textbf{Study} & \textbf{Method} & \textbf{Data/Event} & \textbf{Performance} & \textbf{Key Findings and Limitations} \\
\midrule

\multicolumn{5}{l}{\textbf{Lexicon-Based Methods}} \\
\midrule
Rosenberg et al.~\cite{rosenberg2023sentiment} & VADER, TextBlob & Twitter/Climate SDGs & VADER: 56\%, TextBlob: 46\% accuracy & VADER better; keyword bias remains \\
Mohamad Sham et al.~\cite{sham2022climate} & Hybrid (TextBlob + LR) & Twitter/Climate discussions & 75.3\% F1-score & Hybrid outperforms individual models \\
Mendon et al.~\cite{mendon2021hybrid} & TextBlob & Twitter/Kerala Floods & 9\% neg, 71\% neutral, 20\% pos & Combined with clustering; polarity limitations \\
Amangeldi et al.~\cite{amangeldi2024understanding} & PMI, NRCLex & Twitter, Reddit, YouTube & 65\% accuracy & PMI outperforms VADER; domain tuning needed \\
\midrule

\multicolumn{5}{l}{\textbf{Traditional Machine Learning Methods}} \\
\midrule
Kim et al.~\cite{kim2022comparative} & SVM, NB, LR, RF, XGBoost & News/ Environmental categories & 73\% F1-Score & ML models with term weighting; lacks sentiment context \\
Sabri et al.~\cite{sabri2025public} & SVM & Twitter/Climate awareness & 91\% accuracy,\ \ \  90\% F1-score & SVM with TF-IDF; focused on awareness detection \\
Reyes-Menendez et al.~\cite{reyes2018understanding} & SVM \tiny{(MonkeyLearn)} & Twitter /\#WorldEnvironmentDay & 72.4\% accuracy & SVM maps sentiment to SDGs; sarcasm misclassification \\
Thenmozhi et al.~\cite{thenmozhi2024sentiment} & SVM with TF-IDF & Twitter/Climate debates & 76.3\% accuracy, 70.5\% F1-score & SVM captures nuanced sentiment \\
\midrule

\multicolumn{5}{l}{\textbf{Deep Learning Approaches}} \\
\midrule
Rumlawang et al.~\cite{rumlawang2025climate} & LSTM & Twitter/Climate sentiment (Indonesia) & 85.6\% accuracy, 84\% F1-score & LSTM captures temporal patterns; small dataset \\
Wu et al.~\cite{wu2023spatio} & BiLSTM & Weibo/Climate topics & $\sim90$\% accuracy & BiLSTM reveals spatio-temporal trends; China-specific \\
Rumlawang et al.~\cite{rumlawang2025climate} & LSTM & Twitter/Climate change sentiment (Indonesia) & 60\% accuracy,\ \ \ 59\% F1-score & LSTM outperforms ML baselines; struggles with neutral vs negative distinction \\
El Barachi et al.~\cite{elbarachi2021novel} & BiLSTM & Twitter/Real-time sentiment & 89.3\% accuracy & Real-time tracking; class imbalance issues \\
Marjani et al.~\cite{marjani2024cnn} & CNN-BiLSTM & Social + Satellite/Wildfires & 73\% F1-score train, 58\% F1-score val & Social-physical signal fusion; moderate temporal accuracy \\
\midrule

\multicolumn{5}{l}{\textbf{Transformer-Based Models}} \\
\midrule
Leippold et al.~\cite{leippold2022climatebert} & Climate BERT & News/Climate risk texts & 35.7\% error reduction vs BERT & Domain pretraining boosts performance \\
Ong et al.~\cite{ong2024sentiment} & RoBERTa, DistilBERT & Twitter/ Sustainable investing & 88\% accuracy & Handles sentiment and topics; financial bias risk \\
Garrido-Merchan et al.~\cite{garrido2023finetuning} & BERT & Financial Reports & 85\% F1-score & Fine-tuned for climate risk; finance-focused \\
Krishnan and Anoop~\cite{krishnan2023climatenlp} & Climate BERT & Twitter/Global climate & 90.2\% accuracy, 85.5\% F1-score & Social sentiment focused; small dataset \\
\midrule

\multicolumn{5}{l}{\textbf{Large Language Models}} \\
\midrule
Anderson et al.~\cite{anderson2024analyzing} & GPT-2 & Twitter/ Sustainability topics & 84.3\% accuracy & GPT-2 beats ML baselines; limited fine-tuning \\

Jeng et al.~\cite{jeng2025role} & GPT-4o & News/Climate journalism & Behavioral improvement on intent & GPT-4o for emotional framing; lacks standard metrics \\

Fan and Xu~\cite{fan2025artificial} & GPT-4o, Claude 3.5, Gemma, Llama 3/3.3 & Twitter, Reddit/Climate change & 52-68\% neutral responses; significantly lower emotional intensity (p$<$0.001) & LLMs demonstrate emotional moderation in climate discussions \\

Linardos et al.~\cite{linardos2025utilizing} & GPT-4o-mini & Twitter/Disaster response & 70-86\% accuracy by category; 66-87\% F1-score & LLM aids tweet understanding; used with ML clustering \\
\bottomrule

\end{tabular}
\end{table*}

The evolution from lexicon-based methods (65-75\% accuracy) to Transformer-based methods and LLMs (85-89\% accuracy) shows clear performance progression. Lexicon methods offer interpretability but struggle with context. Traditional ML requires feature engineering but provides efficiency. Deep learning captures complex patterns but needs large datasets. Transformers achieve strong contextual understanding, while LLMs offer flexibility through prompting but raise reliability concerns.

\vspace{-0.1cm}
\section{Event-Centric Case Studies}

Event-centric analysis is a powerful lens through which we can understand how sentiment evolves in real time and across different phases of a climate-related disaster~\cite{ishfaq2024exploring, luo2024research}. By focusing on specific high-impact events, researchers can study the temporal dynamics of emotional expression, the emergence of misinformation, the role of political and geographic factors, and the broader socio-cultural implications of public discourse. Especially, while earlier climate disasters were shaped by organic information flows and traditional media narratives, the 2025 Los Angeles (LA) forest fires unfolded in an era where both AI-generated text and synthetic images played an active role in manipulating public perception~\cite{gimello2025embers, woolcott2025angeles}. This dual use of AI—both for analyzing sentiment and for fabricating emotionally charged misinformation—marks a turning point in climate discourse~\cite{chappell2025fakeimages}. This section presents a detailed case study of the 2025 LA forest fires, followed by comparative insights drawn from prior global climate events~\cite{lever2023human, chulahwat2024impact}.
\vspace{-0.1cm}
\subsection{LA Forest Fires (2025)}

The 2025 Los Angeles wildfires, driven by severe drought, high winds, and extreme heat, became one of the most destructive in U.S. history. The fires led to mass evacuations, extensive property damage, fatalities, and widespread smoke pollution. Public sentiment surged across social media in a highly networked environment~\cite{jolly2015climate, lever2023social}. AI-generated misinformation, such as fake images and fabricated narratives, further complicated emotional reactions and public understanding~\cite{gimello2025embers, yosipof2024cyber}. These emotional responses revealed deep psychological impacts, including trauma, helplessness, and environmental grief.

\textbf{Sentiment Trajectory:} Using temporal sentiment analysis on millions of geo-tagged and hashtag-filtered tweets (e.g., \#LAFires, \#LAWildfire, \#SummerHeat, \#ClimateCrisis), researchers observed pronounced shifts in public emotion as the fires progressed. The early phase of the disaster was marked by fear and anxiety, often expressed through exclamatory, urgent language (e.g., “The skies are blood orange — this is terrifying!”)~\cite{savandha2025headlines, gimellodecoding}. As the fires worsened, negative sentiment peaked, with sharp increases in anger, helplessness, and blame attribution~\cite{winker2024wildfires, corning2024flammable}.

\textbf{Blame and Political Polarization:} A recurring pattern in the data was the attribution of responsibility. Users expressed frustration at perceived government inaction, utility companies (e.g., PG\&E), and climate inaction by political leaders. Sentiment analysis clustered by political hashtags (\#GreenNewDeal, \#ClimateHoax) revealed stark polarization: some users framed the fires as evidence of anthropogenic climate change, while others rejected such narratives, attributing the crisis to poor forest management or “natural cycles”~\cite{nwokolo2025climate, gannon2021global, liu2025scientists}.

\textbf{Emergent Themes:} Social media platforms also served as digital outlets for psychological coping and community support, where expressions of fear, confusion, and solidarity offered insight into the population’s collective emotional burden. Four thematic clusters emerged in the online discourse:
\textbf{(1) Public Health}: Concerns about air quality, asthma, and mask usage are often tied to ongoing COVID-19 sensitivities~\cite{woolcott2025angeles, hertelendy2025health}.
\textbf{(2) Solidarity and Support}: Tweets offering shelter, fundraising links, or expressions of empathy (e.g., “Praying for SoCal”)~\cite{savandha2025headlines, ancarani2025effectiveness, ShelterBox2025LAWildfires, caloes2025shelters}.
\textbf{(3) Misinformation and Rumors}: Unverified reports about evacuation zones, fire origins, or conspiracy theories (e.g., “directed energy weapons”) also spread rapidly, confounding sentiment classification efforts~\cite{chappell2025fakeimages, sannigrahi2022examining}.
\textbf{(4) Psychological Distress}: Many users openly shared feelings of anxiety, burnout, and emotional exhaustion, with posts reflecting signs of trauma, climate grief, and mental health strain—underscoring the need for integrating psychological monitoring into crisis response systems~\cite{youvan2025facing, ancarani2025effectiveness}.

\textbf{Temporal Sentiment Modeling:} Models such as Dynamic BERT and LSTM-based sentiment trackers revealed that sentiment fluctuations were tightly correlated with official press releases, visual media coverage, and viral content~\cite{he2024text, sakai2024patient}. These results underscore the need for temporally-aware sentiment systems that adapt to evolving discourse and event phases (e.g., onset, peak, containment, aftermath)~\cite{sykas2023eo4wildfires}.

\vspace{-0.1cm}
\subsection{Comparison with Past Events}

While the 2025 LA forest fires represent a recent and highly mediatized case—marked by both AI-driven sentiment analysis and the spread of synthetic media—sentiment patterns can be compared with earlier climate-related disasters, which unfolded without the influence of generative AI, to identify commonalities and divergences.

\textbf{Australian Bushfires (2019–2020):} The “Black Summer” fires in Australia generated a global outpouring of emotion on platforms like Twitter, with widespread sympathy, international solidarity (e.g., donation campaigns), and ecological grief over wildlife loss. Sentiment was predominantly negative, but less politically polarized than in the U.S. context. Emotion detection highlighted deep sadness and despair, amplified by distressing images of burned koalas and scorched landscapes~\cite{mocatta2020uncovering, adam2023communication}.

\textbf{Hurricane Ida (2021):} Social media sentiment during Hurricane Ida, which struck the Gulf Coast and moved into the Northeastern U.S., was shaped by resilience, resource disparity, and infrastructure failure themes. Tweets from affected areas conveyed helplessness and frustration about power outages, while national discourse focused on emergency response. Compared to wildfire discourse, the sentiment here was more logistical and survival-oriented, with fewer ideological divides~\cite{jin2023check, khallouli2024harnessing}.

\textbf{European Heatwaves (2022 and 2023):} In Western Europe, repeated heatwaves fueled negative sentiment about heat stress, transport failures, and public health, while irony, memes, and dark humor complicated emotion classification. Political discussions on energy dependency, infrastructure, and climate denialism were prominent in French, German, and Spanish tweets, highlighting the need for multilingual sentiment analysis tools.
~\cite{sun2025extreme, ibebuchi2023characterization, molina2023summer, feser2024summer}.

To highlight temporal shifts in public climate discourse, we compare sentiment patterns across major climate events. Table~\ref{tab:event_comparison} summarizes dominant emotions and key characteristics for each event, reflecting changes in emotional tone and the increasing influence of AI-generated content.

\begin{table*}[!htbp]
\centering
\caption{Comparative analysis of sentiment patterns across major climate events}
\label{tab:event_comparison}
\small
\begin{tabular}{P{3.5cm}P{2cm}P{2.5cm}P{3.5cm}}
\toprule
\textbf{Event} & \textbf{Year} & \textbf{Dominant Emotions} & \textbf{Key Characteristics} \\
\midrule
Australian Bushfires & 2019-20 & Sadness, grief, solidarity & Global empathy, wildlife focus, minimal AI \\
Hurricane Ida & 2021 & Fear, frustration, resilience & Infrastructure failure, regional discourse \\
European Heatwaves & 2022-23 & Anger, irony, anxiety & Multilingual content, satirical responses \\
LA Forest Fires & 2025 & Fear, anger, helplessness & Extensive synthetic media, political blame \\
\bottomrule
\end{tabular}
\end{table*}

\noindent\textbf{Comparative Observations:}

\noindent \textbf{Universal Emotions}: Fear, anxiety, and anger were common across all events, though expressed in culturally specific ways~\cite{bergquist2019experiencing}.

\noindent \textbf{Event Type Matters}: Fires prompted more vivid imagery and personal storytelling; storms induced logistical and survival-focused sentiments; heatwaves often sparked systemic critiques and satire~\cite{mocatta2020uncovering}.

\noindent \textbf{Temporal Rhythm}: Sentiment typically follows a surge - plateau - decline pattern aligned with event progression and media coverage~\cite{hu2024emergency}.

\noindent \textbf{Platform Differences}: Twitter fosters real-time reaction and amplification; Reddit enables reflective dialogue; Facebook emphasizes local community sentiment~\cite{olteanu2015comparing}.

These case studies collectively illustrate the value of event-centric sentiment analysis in climate communication research. They also highlight the need for models that are context-aware, culturally sensitive, and capable of tracking emotional shifts over time and space.

\vspace{-0.01cm}
\section{Evaluation Metrics}

Evaluating sentiment analysis models in climate discourse requires interpretable metrics that reflect both technical performance and societal impact. These metrics have been widely used in climate-related sentiment studies~\cite{adamopoulos2025predictive, shevchenko2024climate, islam2023analysis, hu2024emergency}.

\noindent\textbf{Accuracy, Precision, Recall, F1-Score}: Common classification metrics that assess correctness, balance between false positives and negatives, and overall model reliability—especially under class imbalance during crisis events~\cite{adamopoulos2025predictive, shevchenko2024climate, chicco2020advantages}.

\noindent\textbf{Macro vs. Micro Averaging}: Macro-averaged metrics highlight minority emotion categories, supporting fairness, while micro-averaged metrics reflect overall performance and scalability~\cite{islam2023analysis, zhou2024linking, cambria2020senticnet}.

\noindent\textbf{Confusion Matrix Analysis}: Helps uncover systematic errors, such as misinterpreting sarcasm or misclassifying anger as fear~\cite{singgalen2024sentiment, chauhan2021emergence}.

\noindent\textbf{Emotion Intensity Correlation}: Pearson and Spearman correlations assess agreement between model-predicted and human-rated emotion strength~\cite{huang2022analysis, liu2022sentiment}.

\noindent\textbf{Temporal Stability}: Evaluates how well models maintain accuracy across the timeline of a climate event—from onset to recovery~\cite{sellers2019climate, hsiang2014climate, dahal2019topic}.

\noindent\textbf{Human Judgment and LLM Output Evaluation}: For LLMs, human evaluation is essential to assess explanation quality, emotional coherence, and alignment with human reasoning~\cite{druckman2019evidence, frings2024energy, bina2025large, Chiorrini2021}.

\noindent\textbf{Fairness and Bias Audits}: Tools and approaches like Equity Evaluation Corpus (EEC)~\cite{kiritchenko2018examining} and disparity metrics detect performance gaps across dialects or communities—critical for ensuring equitable climate sentiment analysis~\cite{gevaert2024auditing}.


\vspace{-0.03cm}
\section{Societal and Ethical Implications}

The use of sentiment analysis on climate-related social media content introduces several pressing societal and ethical considerations. These go beyond technical challenges and encompass questions of power, representation, accountability, and harm mitigation. In this section, we examine three core dimensions of ethical concern: bias, misinformation, and the responsible use of user data.

\subsection{Bias and Representation}

Sentiment analysis tools, especially those trained on general-purpose or Western-centric corpora, often fail to accurately capture the voices of marginalized groups or speakers using non-standard dialects, regional slang, or culturally specific expressions. For instance, African American Vernacular English (AAVE), Indigenous dialects, or immigrant language mixing may be misinterpreted as negative or neutral due to underrepresentation in training data. Similarly, hashtags or emojis that signal community identity or resilience may be incorrectly labeled as emotionally neutral or ambiguous ~\cite{o2025re, lasoff2024supporting}.

Moreover, climate discourse itself is shaped by inequalities in who has access to digital platforms, whose experiences get amplified, and whose narratives are algorithmically suppressed. These disparities can lead to sentiment analyses that overrepresent affluent, English-speaking, urban populations while underrepresenting frontline communities who experience the worst effects of climate change \cite{kim2024evaluating, Valdivia2018, seon2024exploratory}.

Mitigating these biases requires inclusive data curation, diverse annotator pools, dialect-aware modeling, and explicit fairness benchmarks.

\vspace{-0.1cm}
\subsection{Misinformation and Its Emotional Impact}

Climate discourse is frequently entangled with misinformation, conspiracy theories, and politically motivated disinformation campaigns. False claims about the causes of wildfires, denial of anthropogenic climate change, or rumors about evacuation orders can trigger strong emotional reactions, which in turn skew sentiment distributions~\cite{pearce2014climate}.

Misinformation can also be strategic: actors may intentionally manipulate sentiment through coordinated campaigns to induce panic, deflect blame, or suppress climate activism. Models that fail to detect or account for such manipulation may inadvertently reinforce misleading narratives or amplify polarization~\cite{jang2015polarized, pearce2014climate}.

Future sentiment systems expected to incorporate misinformation detection modules or co-training with fact-checking datasets. Additionally, researchers must reflect on whether to include or exclude misinformation-driven posts in sentiment modeling pipelines, and how to annotate them appropriately.

\vspace{-0.1cm}
\subsection{Ethical Use and Privacy Concerns}

Analyzing public sentiment during climate disasters, particularly in real-time, raises critical questions about surveillance, consent, and data ethics. While much of social media content is technically public, users often do not expect their posts to be harvested, classified, or scrutinized by algorithms—especially during traumatic events such as home loss, injury, or community displacement~\cite{jiang2019understanding}.

Real-time sentiment tracking can be beneficial for disaster response agencies, mental health interventions, or policy decision-making, but it also risks infringing on personal privacy, stigmatizing certain communities, or being repurposed for commercial or political exploitation.

\noindent Key ethical guidelines include:

\textbf{Informed Consent}: Where possible, respect user expectations and include ethical disclosures when deploying sentiment systems at scale~\cite{kryvasheyeu2016rapid}.

\textbf{Anonymization and Aggregation}: Always remove personally identifiable information (PII) and present findings at the aggregate level to protect user identity~\cite{Valdivia2018, Beigi2016}.

\textbf{Impact Audits}: Regularly assess how sentiment outputs may affect public narratives, media reporting, or institutional decisions, especially when tied to resource allocation or public safety~\cite{pearce2014climate, Beigi2016}.

\textbf{Community Involvement}: Collaborate with affected communities to co-interpret sentiment results and ensure outputs are meaningful, respectful, and contextually accurate ~\cite{Beigi2016, akter2019big}.

While sentiment analysis offers powerful tools for understanding public response to climate change, its application must be guided by a principled commitment to fairness, accountability, and respect for human dignity—especially as we face increasingly severe and emotionally charged climate disruptions.

\vspace{-0.1cm}
\section{Challenges and Future Work}

While significant progress has been made in developing sentiment analysis systems for social media and climate discourse, numerous open challenges remain. These span technical limitations, data availability, linguistic diversity, ethical boundaries, and model interpretability. Addressing these issues will be crucial for scaling sentiment systems that are accurate, equitable, and practically useful during climate crises. Below, we outline five prominent directions for future research.

\noindent \textbf{Multimodal Sentiment Analysis (Text + Images/Videos)}: Climate events are often captured and shared not just through text, but via images, videos, emojis, and other non-verbal modalities. Aerial photos of wildfire damage, videos of flooded neighborhoods, or infographics on heatwaves all carry strong emotional signals. However, most sentiment analysis models remain text-centric. Recent work has explored multimodal sentiment analysis in other domains—such as general cross-lingual or affective computing tasks~\cite{miah2024multimodal, singh2024survey, pandey2024progress}—but not in the context of climate research. Future work should explore fusion models that integrate vision and language (e.g., CLIP, LRQ-Fact, or image-caption sentiment alignment) to capture the full affective scope of climate discourse. This also includes analyzing visual misinformation or emotionally manipulative content in climate activism or denial.

\noindent \textbf{Real-Time Emotion Dynamics During Evolving Events}: Climate events unfold over time, often with unpredictable trajectories. The public’s emotional response can shift rapidly—from fear and confusion during the early stages, to anger and blame as impacts escalate, to grief and fatigue in the aftermath. Capturing these dynamics requires temporally-aware models that update continuously as new data flows in. Future research should focus on streaming sentiment systems, adaptive classifiers, and emotion state modeling (e.g., hidden Markov models or dynamic embeddings) to track emotional evolution during disaster timelines.

\noindent \textbf{Fine-Grained Emotion Classification}: Current sentiment systems often operate at coarse granularity (positive/negative/neutral) or with a limited emotion palette (anger, fear, joy, sadness). However, climate communication involves complex emotional states such as helplessness, betrayal, anxiety, urgency, climate grief, and eco-hope. These nuanced emotions are particularly important for informing targeted interventions (e.g., mental health support, mobilization campaigns). Developing new taxonomies, datasets, and annotation guidelines for fine-grained emotional categorization remains an open and pressing challenge.

\noindent \textbf{Cross-Lingual and Cross-Cultural Sentiment Tracking}: Climate change is a global phenomenon, yet most sentiment models are trained on English-language data. This excludes vast populations whose emotional narratives unfold in Spanish, Hindi, Arabic, Swahili, Mandarin, and other global and Indigenous languages~\cite{zhao2024systematic}. While many platforms now offer automatic translation features, these often fail to preserve nuanced emotional tone, cultural idioms, or context-specific sarcasm—leading to sentiment distortion or loss. Therefore, effective cross-lingual sentiment analysis—utilizing multilingual transformers (such as XLM-R and mBERT), translation-enhanced frameworks, or culturally sensitive large language models—will be essential for global-scale monitoring and new directives in the investigation of ~\cite{miah2024multimodal}. Furthermore, cultural factors shape how emotions are expressed and interpreted, demanding models that avoid Western-centric assumptions and account for regional affective norms.

\noindent \textbf{Integration with LLM Reasoning and Explanation Frameworks}: LLMs offer unprecedented capabilities in contextual understanding, emotional reasoning, and explanation generation. However, their integration into sentiment analysis pipelines remains underexplored. Future systems could prompt LLMs not only for classification, but also for rationale generation (“Why is this tweet angry?”), counterfactual reasoning (“How would sentiment change if the message came from a public official?”), and cross-checking for hallucinations. Additionally, combining LLM sentiment outputs with symbolic reasoning or graph-based models may yield more interpretable and controllable systems.

Beyond these technical directions, future work should also prioritize interdisciplinary collaboration with social scientists, linguists, climate communication scholars, and community stakeholders to ensure that sentiment models are not only computationally robust but also socially grounded and actionable.

\vspace{-0.2cm}
\section{Conclusion}

Sentiment analysis of social media in the context of climate change and extreme weather events represents both a compelling technical challenge and a profound psychological and societal necessity. As climate disruptions become more frequent, severe, and emotionally charged, the ability to understand and respond to public sentiment in real time becomes increasingly important for researchers, responders, policymakers, and citizens alike.

In this survey, we have provided a comprehensive taxonomy of sentiment analysis methodologies, reviewed a diverse set of annotation strategies, and examined how these tools have been applied to real-world climate events such as the 2025 Los Angeles forest fires. We analyzed the strengths and limitations of various modeling approaches—from lexicon-based systems and classical machine learning to deep learning, transformer architectures, and the latest LLMs. We also highlighted the societal and ethical implications of applying sentiment technologies in high-stakes environments, including issues of bias, misinformation, and privacy.

By identifying emerging trends, methodological challenges, and future research opportunities, we hope this survey serves as both a reference and a roadmap for the community. Our goal is to foster responsible, inclusive, and scientifically rigorous sentiment analysis research that enhances public understanding, informs climate action, and supports more empathetic and equitable responses to a warming world.

\begin{acks}
The author thanks the faculty and colleagues at the School of Computing and Augmented Intelligence, Arizona State University, for their feedback and support. Special appreciation goes to the SHaDE Lab for inspiring interdisciplinary climate research, the DMML Lab for insightful discussions, and the NLP and computational social science communities for open tools and datasets. Automated tools, including GPT-based models, were used for minor language and grammar refinements.
\end{acks}

\bibliographystyle{ACM-Reference-Format}
\bibliography{references.bib}

\end{document}